\newcommand{\diracslash}[1]{#1\llap{/\kern2pt}}

\def\bearr{\begin{eqnarray}}

\def\eearr{\end{eqnarray}}

\newcommand{\be}{\begin{equation}}

\newcommand{\ee}{\end{equation}}

\newcommand{\bea}{\begin{eqnarray}}

\newcommand{\eea}{\end{eqnarray}}

\newcommand{\ba}[1]{\begin{array}{#1}}

\newcommand{\ea}{\end{array}}

\newcommand{\eqrf}[1]{Eq.\ (\ref{#1})}

\newcommand{\eqrftw}[2]{Eqs.\ (\ref{#1}) and (\ref{#2})}

\documentclass[preprint,secnumarabic,floats,amssymb,bibnotes,nofootinbib,tightenlines]{revtex4}

\usepackage{graphicx}
\usepackage{fixltx2e}
\usepackage{amsmath}
\usepackage{amssymb}
\usepackage{tabularx}

\usepackage{latexsym}

\usepackage{epsfig}

\usepackage{subfigure}
\usepackage{hyperref}
\usepackage[outdir=./]{epstopdf}

\begin{document}
\title{Observational constraints on power law Starobinsky inflation}
\author{Saisandri Saini and Akhilesh Nautiyal}
\affiliation{Department of Physics, Malaviya National Insitute of Technology, Jaipur,
JLN  Marg, Jaipur-302017, India}

\begin{abstract}
In this work we revisit  power law, $\frac{1}{M^2}R^\beta$,  inflation to find the  deviations from $R^2$ inflation
allowed by current CMB and LSS observations. We compute the power spectra for scalar and tensor perturbations
 numerically and perform MCMC analysis  to put constraints on parameters $M$ and $\beta$ from Planck-2018, BICEP3 and other
LSS observations. We consider general reheating scenario and also vary the number of e-foldings during inflation, $N_{pivot}$,
 along with the other parameters. We find 
$\beta = 1.966^{+0.035}_{-0.042}$, $M= \left(3.31^{+5}_{-2}\right)\times 10^{-5}$  
and $N_{pivot} = 41^{+10}_{-10}$ with $95\%\, C.\, L.$. This indicates that the current observations allow deviation from 
Starobinsky inflation. The scalar spectral index, $n_s$, and tensor-to-scalar ratio, $r$, derived from these parameters,
are consistent with the Planck and BICEP3 observations.   

\end{abstract}

\maketitle
\section{Introduction} The idea of inflation \cite{Guth:1980zm} was introduced to solve various problems of the Big-Bang theory. 
Later it was realized \cite{Mukhanov:1981xt,Starobinsky:1982ee,Guth:1985ya} that it provides seeds for cosmic microwave
background anisotropy and large scale structure of the universe. During inflation the potential energy of a scalar 
field, named as inflaton, dominates the energy density of the universe, and causes quasi-exponential expansion of the universe 
for a very short period of time. The quantum fluctuations in the inflaton field, which are coupled to the metric fluctuations,
generate primordial density perturbations. There  are also quantum fluctuations in the spacetime geometry during 
inflation responsible for the primordial gravitational waves, also known as tensor perturbations. 
Inflation predicts nearly scale invariant, adiabatic and Gaussian perturbations that are in excellent agreement with 
various CMB observations such as COBE \cite{Smoot:1992td}, WMAP \cite{Komatsu:2010fb}
and Planck \cite{Ade:2015lrj, Planck:2018jri}. 
The choices for inflaton potential are derived from various particle physics models and string 
theory, which provide a large class of inflaton potentials \cite{Martin:2013tda}; however, we lack a unique model of inflation.

The first self-consistent model of inflation was proposed by Starobinsky in 1980 \cite{Starobinsky:1980te}, 
where  inflation is achieved by  $\frac{1}{M^2} R^2$ interaction, $R$ being the Ricci scalar,
 in the Einstein-Hilbert action without additional scalar field. Transforming
to the Einstein frame, the $R^2$ Starobinsky model gives rise to plateau potential of the inflaton field. 
The $R^2$ Starobinsky inflation is of great interest as it is one of the best suited models of inflation from recent 
Planck observations \cite{Planck:2018jri}, and  it also incorporates a graceful exit to the radiation dominated epoch via a 
period of reheating \cite{Vilenkin:1985md,Mijic:1986iv, Ford:1986sy}, 
where the standard model particles are created through the oscillatory decay of the inflaton, 
called as scalaron in case of $R^2$ inflation.

In this work, we investigate  the generalization of Starobinsky inflation by considering a power law,  $\frac{1}{6M^2} \frac{R^{\beta}}{M_{Pl}^{2\beta-2}}$, correction to the Einstein-Hilbert action. 
 Here $\beta \approx 2$ and not necessarily an integer, allowing a small deviation from $\beta=2$. The $R^\beta$ Lagrangian was
first considered in the context of higher order metric theories of gravity \cite{Schmidt:1989zz,Maeda:1988ab} 
and was then applied to inflation \cite{Muller:1989rp,Gottlober:1992rg} as a generalization of $R^2$ inflation 
(see also \cite{DeFelice:2010aj,Martin:2013tda,Nojiri:2010wj,Nojiri:2017ncd} for detailed review). It was shown 
in \cite{Codello:2014sua,Ben-Dayan:2014isa,Rinaldi:2014gua} that $R^\beta$ term, with
$\beta$ slightly different from $2$ can  arise as a quantum correction to Starobinsky $R^2$ term in the Einstein-Hilbert action.  
 It was also shown in \cite{Costa:2014lta} that the models of Higgs field as inflaton with local Weyl symmetry are
 equivalent to generalized Starobinsky inflation in Einstein frame. 
 Power law terms in the Einstein-Hilbert action in the Jordan frame can be reconstructed from various scalar potentials in 
Einstein frame \cite{Sebastiani:2013eqa}.
It has also been shown in \cite{Cai:2014bda} that
the power law Starobinsky inflation can be embedded into a general class T-models \cite{Kallosh:2013hoa}. Modifications to 
$R^2$-term has also been obtained in \cite{Odintsov:2021wjz} by considering quasi-de Sitter evolution in $f(R)$ gravity. 
$R^\beta$ inflation 
became popular in 2014 when BICEP2 reported large value of tensor-to-scalar 
ratio \cite{BICEP2:2014owc}
$r=0.2^{+0.07}_{-0.05}$. It was shown \cite{Codello:2014sua,Martin:2014lra, Costa:2014lta} that this model 
could generate large $r$ compared to $R^2$ inflation  for $\beta$ slightly smaller than $2$. 
However, it was found later that the BICEP2 signal of $B$-mode polarization is not of primordial
origin, but, due to an unknown amplitude of foreground dust emission \cite{Planck:2014dmk}. 
$R^\beta$ Lagrangian was also used to construct a unified model of inflation and dark energy 
in $f(R)$ gravity \cite{Artymowski:2014gea}.  
It is shown in  \cite{Odintsov:2019mlf,Odintsov:2020iui} that the presence of Chern-Simons term along with $R^\beta$
  can significantly reduce the tensor-to-scalar ratio as well as predicts axions as dark matter candidates.   

Another interesting aspect of Starobinsky inflation is that a no-scale supergravity model of inflation with a modulus field 
and the inflaton field with a minimal Wess-Zumino superpotential gives the same $F$-term potential in the Einstein frame as 
the Starobinsky model \cite{Ellis:2013xoa}. It has been shown in \cite{Ellis:2013nxa,Ellis:2018zya} that there are various 
possible scenarios of non-scale supergravity that can reproduce the effective 
potential of Starobinsky model and other related models. The Starobinsky model can also be derived from the $D$-term potential 
in supergravity models of inflation \cite{Buchmuller:2013zfa,Ferrara:2013rsa,Farakos:2013cqa,Ferrara:2013kca}. 
A no-scale supergravity model, with inflaton potential equivalent to power law Starobinsky potential in Einstein frame,
was obtain in \cite{Chakravarty:2014yda} by using a $(\Phi+\bar\Phi)^n$ term in the no-scale K\^ahler potential with 
Wess-Zumino form of the superpotential. 

It was also shown in \cite{Chakravarty:2014yda} that a small deviation 
from $\beta=2$ can give the tensor-to-scalar ratio $r\sim\,O(0.1)$. The analysis of \cite{Chakravarty:2014yda} was limited
to $\beta\le 2$ and they used slow-roll approximation to obtain the observational constraint on the model parameters.
The consistency relations among the scalar spectral index, the tensor-to-scalar ratio and the running of scalar spectral
index were derived in  \cite{Motohashi:2014tra} and the observational constraints 
on  $\beta$ were found for various choices of $n_s$, $r$ and $N_k$, again using the slow-roll approximation.
However, the parameter $\beta$ (denoted by $p$ in \cite{Motohashi:2014tra}) was varied between $1.80$ and $2.1$ for the
analysis. 
The attractor solutions for $R^\beta$ model in the Jordan and Einstein frame, for $1.9\le \beta \le 2.01$, 
were also studied in \cite{Odintsov:2022bpg}, and  it was shown, using slow-roll conditions, that $R^\beta$ inflation
 is viable in both the frames. Observational constraints on $R^\beta$ inflation are
also obtained in \cite{Meza:2021xuq} numerically integrating the perturbation equations, however, the entire region of the 
parameter space is not explored and only some selected values of $\beta$ are used. The variations from Starobinsky potential
in the Einstein frame has also been studied in \cite{SantosdaCosta:2020dyl} based on a potential derived from brane 
inflation, and it is found that data allows deviation from Starobinsky model.

In our work we use MODECODE \cite{Mortonson:2010er} to explore the parameter space of $R^\beta$ model. In MODECODE the 
background and perturbation equations for inflation are solved numerically without the usual 
slow-roll approximation, and the power spectra for scalar and tensor perturbations are computed. 
These power spectra are used in CAMB \cite{Lewis:1999bs} to compute the angular power spectra for CMB anisotropy and 
polarization, which is then interfaced with COSMOMC \cite{Lewis:2002ah}, which performs the Markov chain Monte Carlo
analysis for parameter estimation. With MODCODE the parameters of inflationary potential can be constrained
directly from the CMB observations; the standard inflationary parameters, like $r, n_s$ are treated as
derived parameters. We vary $\beta$ between $1.9$ to $2.07$ along with $M$ and $N_{pivot}$ to find the 
best fit parameters of the model and to look for any deviation from Starobinsky model, $\beta=2$. 

The paper is organized as follows. In section \ref{powerlaweinsteinframe} we obtain the potential for power law 
Starobinsky model in Einstein frame using conformal transformations. 
In section \ref{inflationpowerlaw} we obtain the equation of motion for scalar field and perturbation equations used in 
MODECODE. In section \ref{observconstra} we compute the CMB power spectra using MODECODE and CAMB and use COSMOMC to put
constraints on the parameters of power law Starobinsky inflation. We summarize our results and give our conclusions in section
\ref{conclusions}.

\section{Power law Starobinsky model} \label{powerlaweinsteinframe}
\begin{figure}[h]
\begin{center}
\includegraphics[width=12cm, height = 7cm]{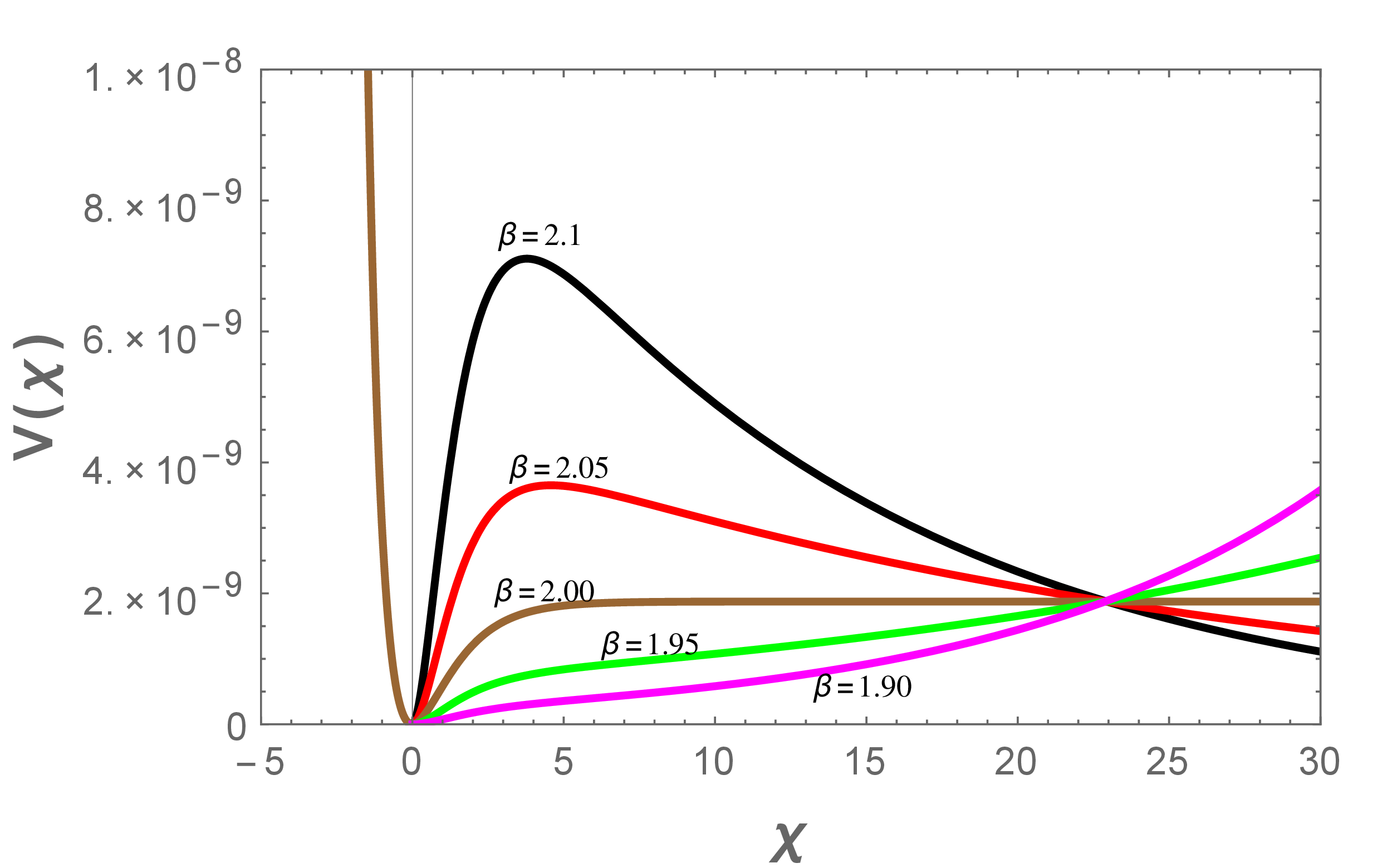}
\caption{The  potential (\ref{pot}) for various values of  $\beta$. The value of M is fixed at $M = 5\times 10^{-5}$, 
and the values of potential and scalar field are in $M_p = 1$ units.}
\label{fig:powerplot}
\end{center}
\end{figure}

The power law Starobinsky inflation is a special case of $f(R)$ gravity, where the   action  is given as \cite{DeFelice:2010aj, Nojiri:2010wj}
\begin{equation}
S_J = \frac{-M_{Pl}^2}{2}\int \sqrt{-g} f(R) d^4 x. \label{SJ} 
\end{equation}
For power law Starobinsky inflation the function $f(R)$ has the form \cite{Chakravarty:2014yda} 
\begin{equation}
 f(R) = \left(R+\frac{1}{6M^2}\frac{R^{\beta}}{M_{Pl}^{2\beta-2}}\right), \label{frpowerlaw}
\end{equation}
where $M_{Pl}^2 = (8\pi G)^{-1}$, g is the determinant of the metric $g_{\mu \nu}$, and $M$ is a dimensionless real 
parameter. The subscript J in \eqrf{SJ} stands for Jordan frame, where the action is a non-linear function of the Ricci 
scalar.  We can rewrite \eqrf{SJ} as
\begin{equation}
S_J = \int d^4x\sqrt-g\left(\frac{-M_{Pl}^2}{2}FR+U\right), \label{Ufr} 
\end{equation}
where
\begin{equation}
 U = \frac{(FR-f)M_{Pl}^2}{2}.  
\end{equation}
Here, $F$ is the first derivative of $f(R)$ with respect to $R$. The action in the Einstein frame can be obtained with the 
conformal transformation $\tilde{g}_{\mu \nu}(x)=\Omega(x)g_{\mu \nu}(x)$, where $\Omega$ is the conformal factor and a tilde 
represents quantities in the Einstein frame. \\
The Ricci scalar $R$ in the Jordan frame is related to the Ricci scalar $\tilde{R}$ in the Einstein frame as:
\begin{equation}
R = \Omega \left(\tilde{R} + 3\widetilde{\Box}\omega - \frac{3}{2}\tilde{g}^{\mu \nu}\partial_{\mu}{\omega}\partial_{\nu}{\omega}\right),\label{ricci}
\end{equation}
where $\omega \equiv \ln\Omega$, $\widetilde{\Box}\omega \equiv \frac{1}{\sqrt{-\tilde{g}}}\partial_{\mu}\left(\sqrt{-\tilde{g}}\tilde{g}^{\mu\nu}\partial_{\nu}{\omega}\right)$ and  $\partial_{\mu}{\omega} = \frac{\partial \omega}{\partial \tilde{x}^{\mu}}$.
We choose  $\Omega = F$ to obtain the action in the Einstein frame and also introduce a new scalar field $\chi$ 
defined by 
\be
\chi \equiv \sqrt{\frac{3}{2}} M_{Pl}\ln F. \label{fchi} 
\ee
This gives $\Omega = \exp\left(\frac{2\chi}{\sqrt{6}M_{Pl}}\right)$.
The action \eqrf{SJ} gets transformed to an Einstein Hilbert form using \eqrf{ricci} and 
relation$\sqrt{-g}=\Omega^{-2}\sqrt{-\tilde{g}}$ as 
\begin{equation}
S_E = \int d^4x\sqrt{-\tilde{g}}\left(\frac{-M_{Pl}^2}{2}\tilde{R}+\frac{1}{2}\tilde{g}^{\mu \nu}\partial_{\mu}{\chi}\partial_{\nu}{\chi}+V(\chi)\right),\label{SE}
\end{equation}
where $V(\chi)$ is the Einstein frame potential given by
\begin{equation}
V(\chi) = \frac{(RF(R)-f(R))M_{Pl}^2}{2F(R)^2}, \label{potfr}
\end{equation}
The potential (\ref{potfr}) for power law Starobinsky model (\ref{frpowerlaw})  in the Einstein frame becomes
\begin{equation}
V({\chi}) = (\frac{{\beta}-1}{2})\left(\frac{6M^2}{{\beta}^{\beta}}\right)^{\frac{1}{\beta-1}} \exp\left[{\frac{2\chi}{\sqrt 6}\left(\frac{2-\beta}{\beta-1}\right)}\right]\times \left(1-\exp\left(\frac{-2\chi}{\sqrt 6}\right)\right)^\frac{\beta}{\beta-1}. \\ \label{pot}
\end{equation}
where we have taken $M_{Pl} = 1$ and we will use this from now on. We can also see that, for $\beta=2$, 
the potential \ref{pot} reduces to Starobinsky $R^2$ inflation. The potential (\ref{pot}) for various choices of $\beta$ around
$\beta=2$ is depicted in Fig.~\ref{fig:powerplot}. The slow-roll inflation occurs in the regime where $\chi >0$. The potential
is flat for $\beta=2$ for large values of $\chi$ and it asymptotically approaches a constant value. However, the
potential  gets steeper for $\beta<2$, which gives larger tensor-to-scalar ratio as compared to $\beta=2$. In case of 
$\beta>2$ the potential first increases with $\chi$ and attains a maximum value 
at $\chi= M_{Pl} \sqrt{\frac{3}{2}}\ln\left[\frac{2\left(\beta-1\right)}{\beta-2}\right]\equiv \chi_m$, then it decreases and goes to 
zero for large $\chi$. Thus, the inflation can occur for $\chi$ rolling between $0\le\chi\le\chi_m$ or $\chi > \chi_m$.
We will consider $\chi < \chi_m$ for our analysis to study the deviation from $R^2$ inflation.  
We solve the background  evolution equations and perturbation equations using potential (\ref{pot}) numerically using 
MODECODE. The necessary equations are discussed in the proceeding section.

\section{Inflationary dynamics} \label{inflationpowerlaw}
\subsection{Background Equations} 
We analyze the power law Starobinsky inflation in the Einstein frame.
During inflation the energy density of the scalar field $\chi$ dominates the Universe, and hence
the expansion  is governed by the Friedmann equations:
\bea
H^2 &=& \frac{1}{3M_{Pl}^2}\left[\frac{1}{2}\dot{\chi^2} + V(\chi)\right]. \label{H2}\\
\dot H &=& -\frac{1}{2 M_{Pl}^2} \dot\chi^2 \label{hprime}
\eea
The equation of motion for  $\chi$ is the Klein-Gordon equation in an expanding spacetime: 
\be
\ddot{\chi} + 3H\dot{\chi} + \frac{dV(\chi)}{d\chi} = 0. \label{evo}
\ee
Here, the dot stands for the differentiation with respect to cosmic time. 
Since we choose the number of e-folding,  $N = \ln a $ as the
independent variable to solve mode equations numerically, the background equations for Hubble parameter \eqrftw{H2}{hprime}
and the scalar field $\chi$ \eqrf{evo}  are expressed  in terms of $N$ as
\bea
H^2 &=& \frac{\frac{1}{3M_{Pl}^2}V(\chi)}{1-\frac{1}{6M_{Pl}^2}\chi^{\prime 2}},\label{H2n}\\
H^\prime &=& -\frac{1}{2M_{Pl}^2}H\chi^{\prime 2},\label{Hprimen}
\eea
and 
\be
\chi^{\prime \prime}+\left(\frac{H^\prime}{H}+3\right)\chi^\prime+\frac{1}{H^2}\frac{dV(\chi)}{d\chi} = 0,\label{evon}
\ee
where prime denotes the differentiation with respect to N. These background equations are solved numerically by setting the 
initial conditions such that the field velocity is at its slow-roll value. This makes sure that the (small) initial transient
in the velocity is damped away. The solution is then used as an input to perturbations equations.
\subsection{Perturbation Equation}
The density perturbations generated during inflation are described by gauge invariant comoving curvature perturbations 
$\mathcal{R}$, which is related to Mukhanov-Sasaki variable $u$ as \cite{Mukhanov:1988jd,Sasaki:1986hm} 
\be
u = -z \cal{R}, \label{curvaturepert}
\ee 
where $z=\frac{1}{H}\frac{d\chi}{d\tau}$, $\tau$ denotes the conformal time. The quantity $z$ depends on the background 
evolution and can be determined by solving \eqrftw{Hprimen}{evon}
The evolution equation for Fourier mode $u_k$ in conformal time  is given as
\be
\frac{d^2 u_k}{d\tau^2} + \left(k^2 - \frac{1}{z}\frac{d^2z}{d\tau^2}\right)u_k = 0,\label{uk}
\ee
The primordial power spectrum is defined in terms of the two point correlation function of comoving curvature 
perturbation as
\be
\mathcal{P_\mathcal{R}} = \frac{k^3}{2\pi^2}\langle \mathcal{R}_{k}\mathcal{R}_{k^\prime}^*\rangle\delta^3(k-k^\prime),
\ee
which is related to Mukhanov-Sasaki variable $u_k$ (\ref{curvaturepert}) as\\
\be
\mathcal{P_\mathcal{R}}(k) = \frac{k^3}{2\pi^2}\bigg|\frac{u_k}{z}\bigg|^2, \label{Ps}
\ee
Similarly the mode equation for tensor perturbations generated during inflation is given as
\be
\frac{ d^2v_k}{d\tau^2} + \left(k^2 - \frac{1}{a}\frac{d^2a}{d\tau^2}\right)v_k = 0,\label{vk}
\ee
and the primordial tensor power spectrum is given as
\be
\mathcal{P}_{t}(k) = \frac{4}{\pi^2}\frac{k^3}{M_{Pl}^2}\bigg|\frac{v_k}{a}\bigg|^2.\label{Pt}
\ee
To obtain the scalar and tensor power spectra, the mode equations (\ref{uk}) and (\ref{vk}) are solved numerically.
As we choose e-foldings $N=\ln a$ as independent variables to solve these equations, the background quantity 
$z=\chi^\prime$. Hence \eqrftw{uk}{vk}   can be written  in terms 
of $N$ as
\bea
&u_k^{\prime \prime} + \left(\frac{H^\prime}{H}+1\right)u_k^{\prime}+\Biggl\{\frac{k^2}{a^2H^2}-\left[2-4\frac{H^\prime}{H}\frac{\chi^{\prime \prime}}{\chi^\prime}-2\left(\frac{H^\prime}{H}\right)^2 -5\frac{H^\prime}{H}-\frac{1}{H^2}\frac{d^2V}{d\chi^2}\right] \Biggr\}u_k = 0,\label{ukn}&\\
&v_k^{\prime \prime} + \left(\frac{H^\prime}{H}+1\right)v_k^{\prime} + \left[\frac{k^2}{a^2H^2} -\left(\frac{H^\prime}{H}+2\right) \right]v_k = 0.&\label{vkn}
\eea
The numerical solutions of \eqrftw{ukn}{vkn} are obtained along with the background equations (\ref{Hprimen}), (\ref{evon}) using
Bunch-Davious initial conditions.

The scalar spectral index $n_s$ and the tensor spectral index $n_t$ are determined from the power spectra
obtained numerically using their definitions \cite{Bassett:2005xm}
\bea
n_s &=& 1 + \frac{d \ln\mathcal{P_\mathcal{R}}}{d \ln k},\label{ns}\\
n_t &=& \frac{d \ln\mathcal{P}_{t}}{d \ln k},\label{nt}
\eea
The tensor-to-scalar ratio $r$ is defined by \cite{Bassett:2005xm}
\be
r = \frac{\mathcal{P}_{t}}{\mathcal{P_\mathcal{R}}}.\label{r}
\ee
Planck CMB observations provide constrains on $n_s$ and $r$. However, in our analysis they are derived parameters, 
and the parameters of the inflaton potential (\ref{pot}), $M$ and $\beta$, are directly constrained from the CMB observations.

\section{Observational constraints} \label{observconstra}
To calculate the scalar and tensor power spectrum for the quantum fluctuations generated during inflation, 
we modify the publicly available MODECODE \cite{Mortonson:2010er} for the power law Starobinsky potential (\ref{pot}) in the 
Einstein frame.  
We consider the general reheating scenario, where the  parameter $N_{pivot}$ that represents the 
number of e-foldings from the end of inflation to the time when length scales correspond the Fourier mode
$k_{pivot}$ leave the Hubble radius during inflation, is also varied along with other potential parameters.
ModeCode can be used within CAMB \cite{Lewis:1999bs}. To compute the primordial power spectra at arbitrary values of k in CAMB, 
ModeCode uses cubic spline interpolation on a grid of k values spaced evenly in $\ln k$. CAMB computes the angular power 
spectra for CMB anisotropy and polarization. These CMB power spectra are used in COSMOMC to put 
constraints on the parameters of the inflaton potential along with the other parameters of $\Lambda$CDM model
 from various CMB and large scale structure observations. 
To constrain the parameters $M$ and $\beta$ of inflaton potential (\ref{pot}) we use
Planck-2018 data along with BICEP3 \cite{BICEP:2021xfz}, BAO and Pantheon data.The priors for the  parameters of inflaton potential and 
$N_{pivot}$ are given in Table \ref{Tab:prior}. The priors for the parameter $M$ are sampled logarithmically to cover 
a large range. The parameter $\beta$ is varied around $2$ to consider deviation from Starobinsky inflation. The other parameters
of the $\Lambda$CDM model are also varied along with these three parameters with priors given in \cite{Planck:2018vyg}. 
 For each parameter the MCMC convergence diagnostic tests is preformed over the  four chains using the 
Gelman and Rubin "variance of mean" / "mean of chain variance " $R-1$ statistics.  

\begin{table}[h!]
\label{Tab:prior}
\begin{tabularx}{0.8\textwidth} { 
  | >{\raggedright\arraybackslash}X 
  | >{\centering\arraybackslash}X 
  | >{\raggedleft\arraybackslash}X | }

 \hline
 {\boldmath$log_{10} M $} & $-6.5<log_{10} M<-3.0$  \\
 \hline
 {\boldmath$\beta $} & $1.90<\beta<2.07$    \\
 \hline
 {\boldmath$N_{pivot}$} & $25<N_{pivot}<90$   \\
 \hline
\end{tabularx}
\caption{Priors on model parameters}
\end{table}

The constraints obtained for parameters of potential (\ref{pot}), the e-foldings $N_{pivot}$ and the deriver parameters,
$r$ and $n_s$, are shown in Table \ref{Tab:constraint}. 

\begin{figure}[h!]
\begin{center}
\subfigure[]{
 \includegraphics[width=14cm, height = 4cm]{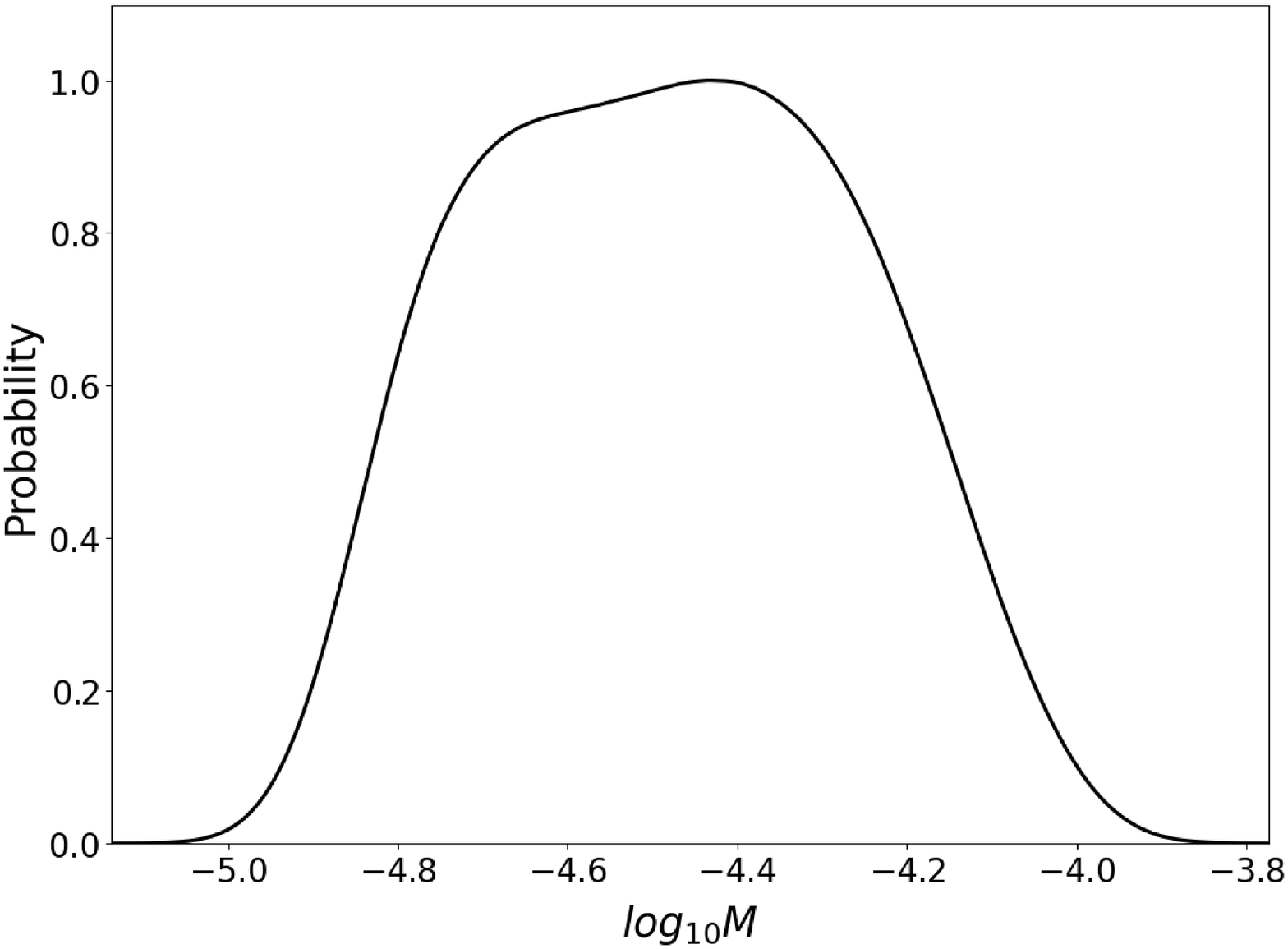}
}
\subfigure[]{
 \includegraphics[width=14cm, height = 4cm]{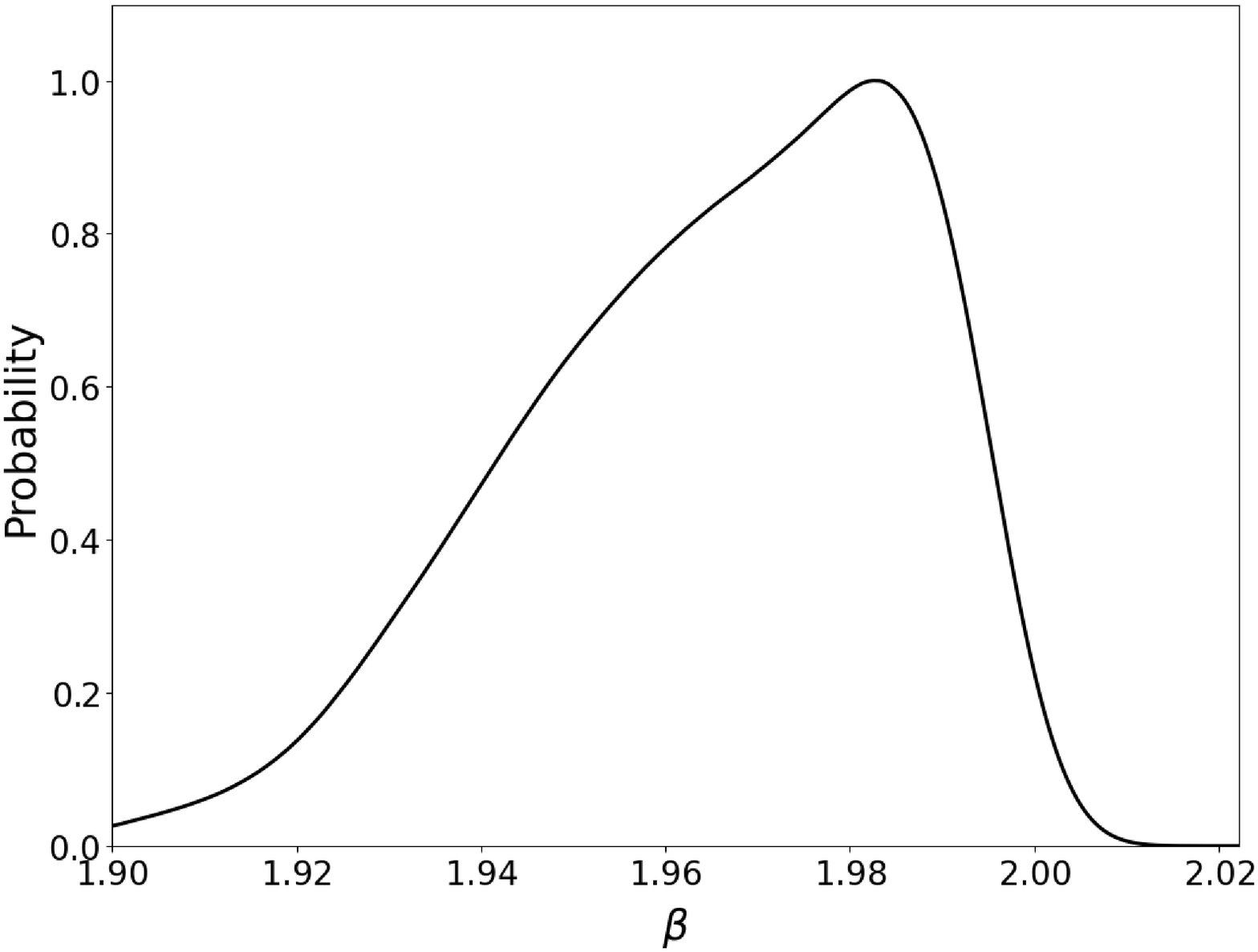}
}
\subfigure[]{
 \includegraphics[width=14cm, height = 4cm]{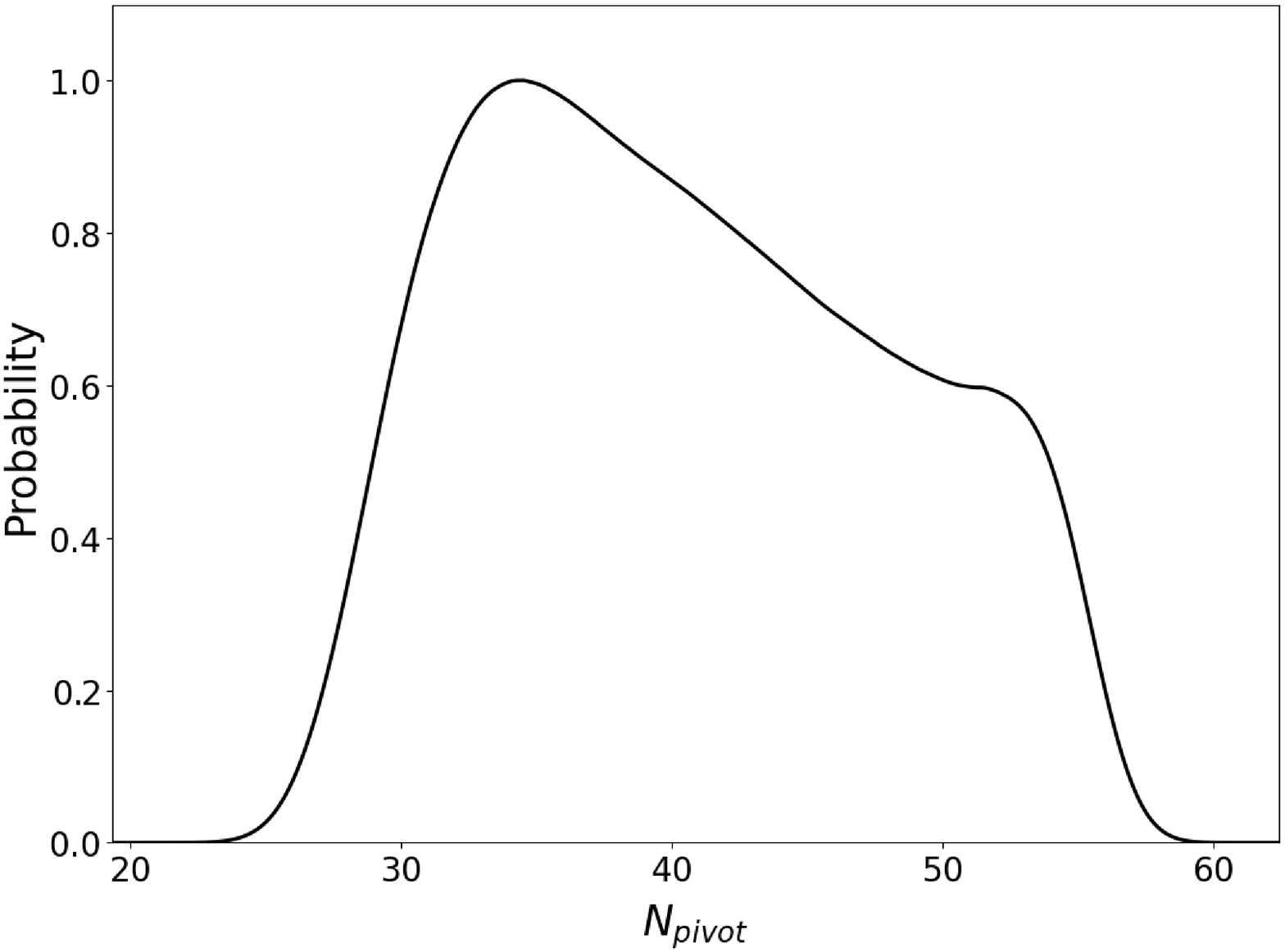}
}
\caption{Marginalized constraints on the potential parameters and $N_{pivot}$ using Planck-2018, BICEP3 and BAO data}
 \label{fig:margconstr}
\end{center}
\end{figure}

\begin{table}[h!]
\begin{tabular} { l  c c c}
  
 Parameter &  68\% limits & 95\% limits   &  99\% limits \\ 
\hline
\hline

{\boldmath$log_{10} M $} & $-4.48^{+0.23}_{-0.26}$ & $-4.48^{+0.40}_{-0.40} $ 
& $-4.48^{+0.50}_{-0.45}$\\

{\boldmath$\beta $} & $1.966^{+0.027}_{-0.015}$ & $1.966^{+0.035}_{-0.042}$ & $1.966^{+0.039}_{-0.056}$ \\

{\boldmath$N_{pivot}$} & $41^{+6}_{-10}$ & $41^{+10}_{-10}$ & $41^{+20}_{-10}$ \\

$n_s $ & $0.9688\pm 0.0036 $ & $0.9688^{+0.0072}_{-0.0071}$ & $0.9688^{+0.0094}_{-0.0094}$\\

$r $ & $0.0198^{+0.0043}_{-0.016}$ & $0.020^{+0.030}_{-0.017}$ 
& $0.020^{+0.048}_{-0.018} $\\
\hline
\hline
\end{tabular}
\caption{ Planck-2018, BICEP3 and BAO constraints on parameters of potential, $r$ and $n_{s}$.}
\label{Tab:constraint}
\end{table}
It is evident from the Table that the best fit value of $\beta$ is
\be
\beta = 1.966^{+0.035}_{-0.042},\, \, \, 95\%\, C.\, L., \label{betadata}
\ee
which indicates that the Planck observations favor deviation from Starobinsky model, $\beta=2$. which lies within 
$2\sigma$ of the best fit value (\ref{betadata}).  
The power law Starobinsky model prefers the number of e-folding
\be
N_{pivot} = 41^{+10}_{-10} ,\, \, \, 95\%\, C.\, L., \label{npivot}
\ee
It can also be seen from the Table \ref{Tab:constraint} that the best fit values of scalar spectral index $n_s$ and 
tensor-to-scalar ratio $r$ derived from the the best fit values of potential parameters, 
for the power law Starobinsky model, is well within the Planck bounds.
The marginalized probability distributions  for various inflationary parameters are shown in Fig.\ref{fig:margconstr}.

\begin{figure}[h!]
\begin{center}
\subfigure[]{
 \includegraphics[width=7cm, height = 6cm]{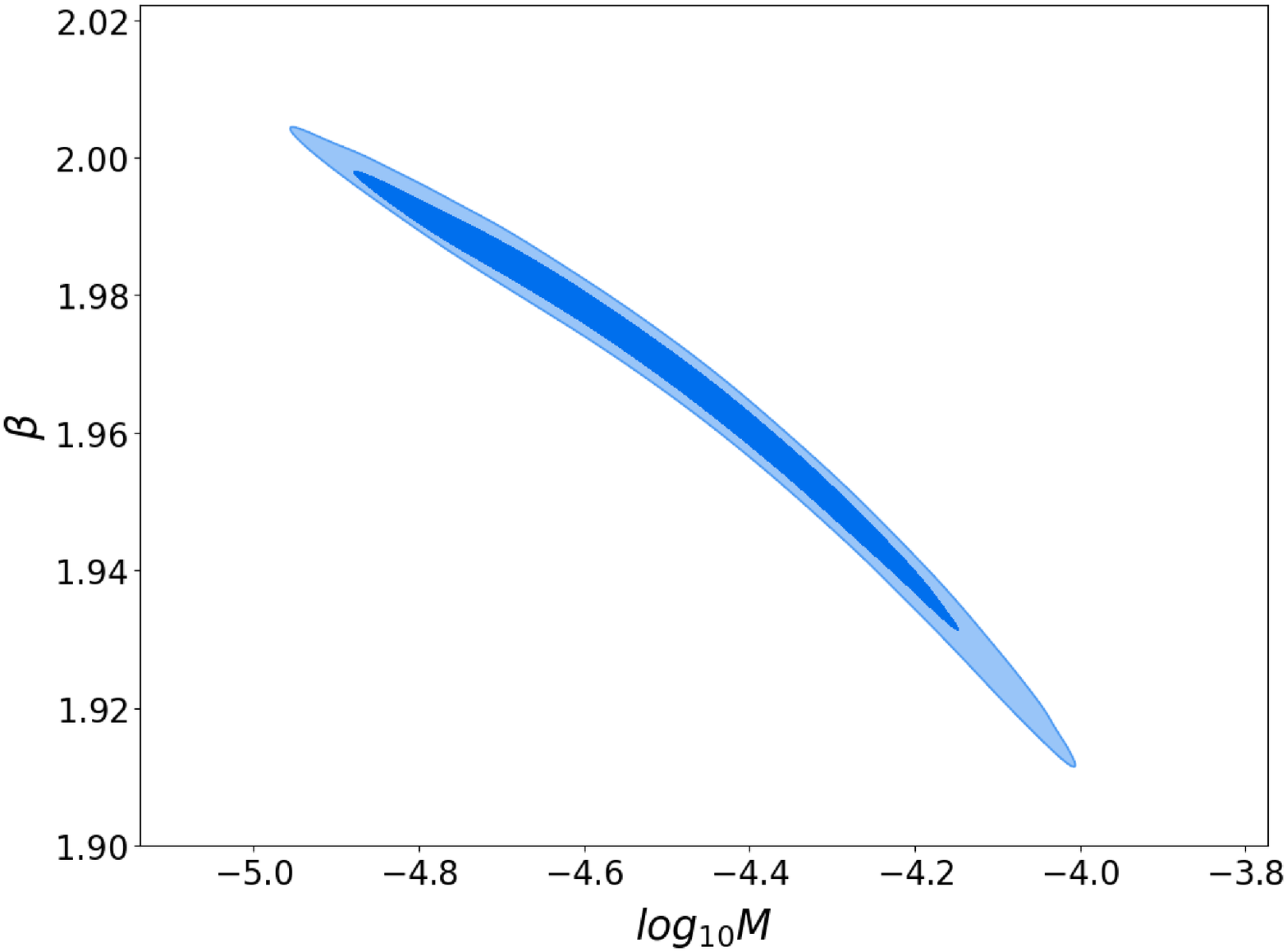}
 \label{fig:betam}
}
\subfigure[ ]{
\includegraphics[width=7cm, height = 6cm]{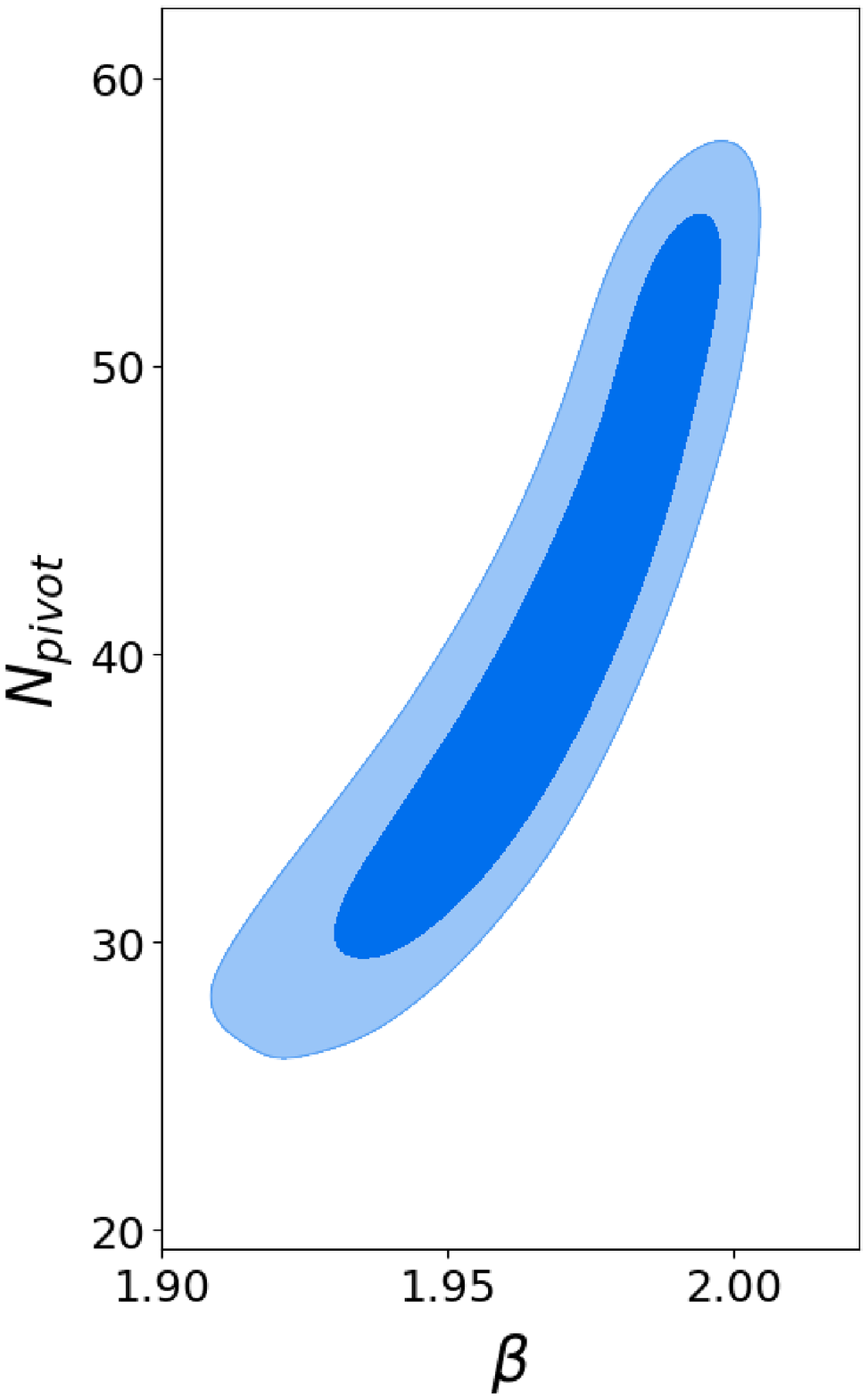}
 \label{fig:betan}
}
\caption{Joint $68\%\, \, C.\, L.$, and $95\%\, \, C.\, L.$ constraints on parameters of potential and 
$N_{pivot}$ using Planck-2018, BICEP3 and BAO data}
\label{fig:betamandn}
\end{center}
\end{figure}

\begin{figure}
\begin{minipage}{0.6\textwidth}
\begin{center}
 \includegraphics[width=8cm, height = 7cm]{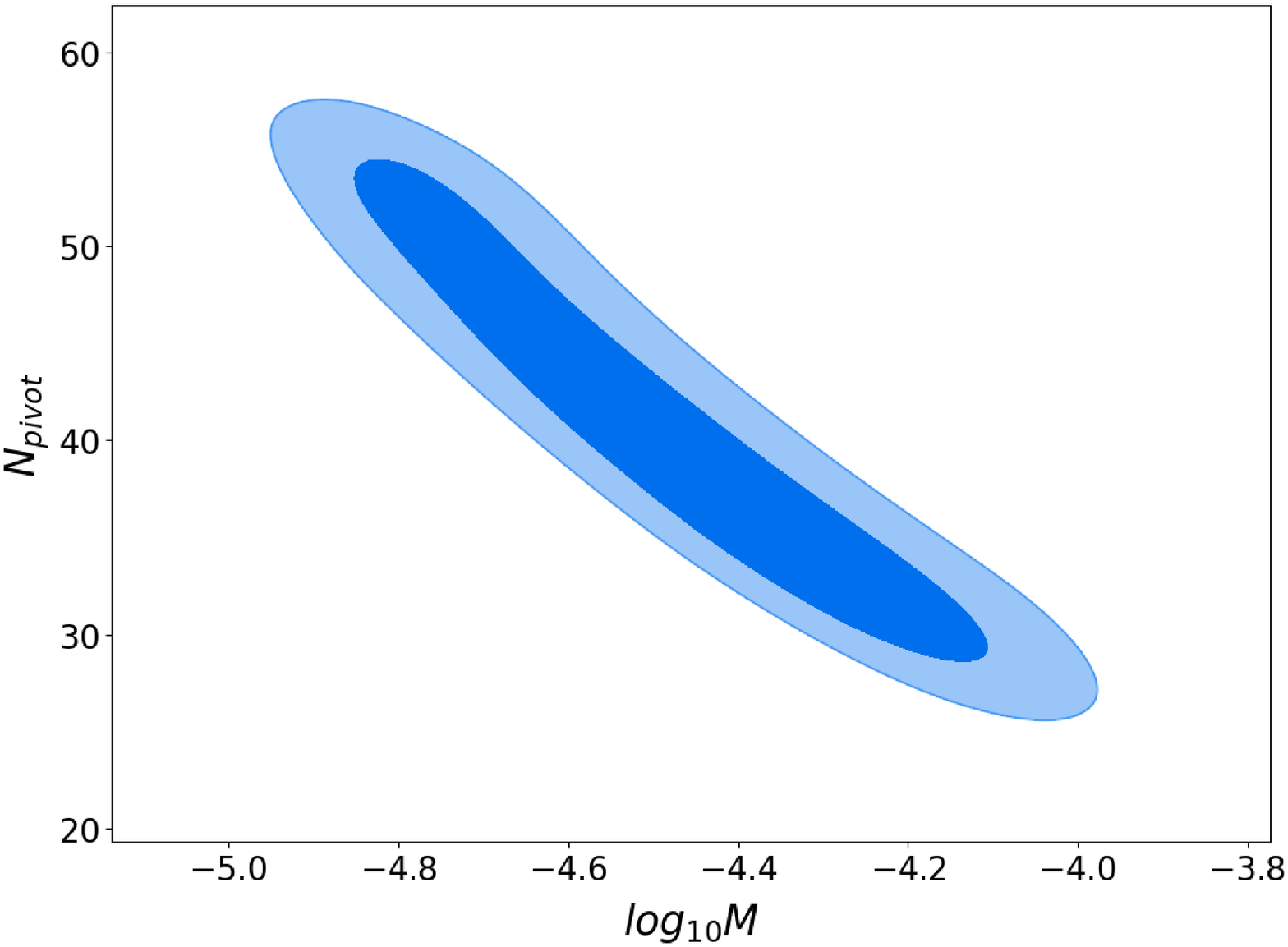}
\caption{Joint  $68\%\, \, C.\, L.$, and $95\%\, \, C.\, L.$ constraints on \\ potential parameter $M$ 
and $N_{pivot}$ from  Planck-2018,\\ BICEP3 and BAO data}
\label{fig:mandn}
\end{center}
\end{minipage}%
\begin{minipage}{.5\textwidth}
\begin{center}
 \includegraphics[width=8.5cm, height = 7cm]{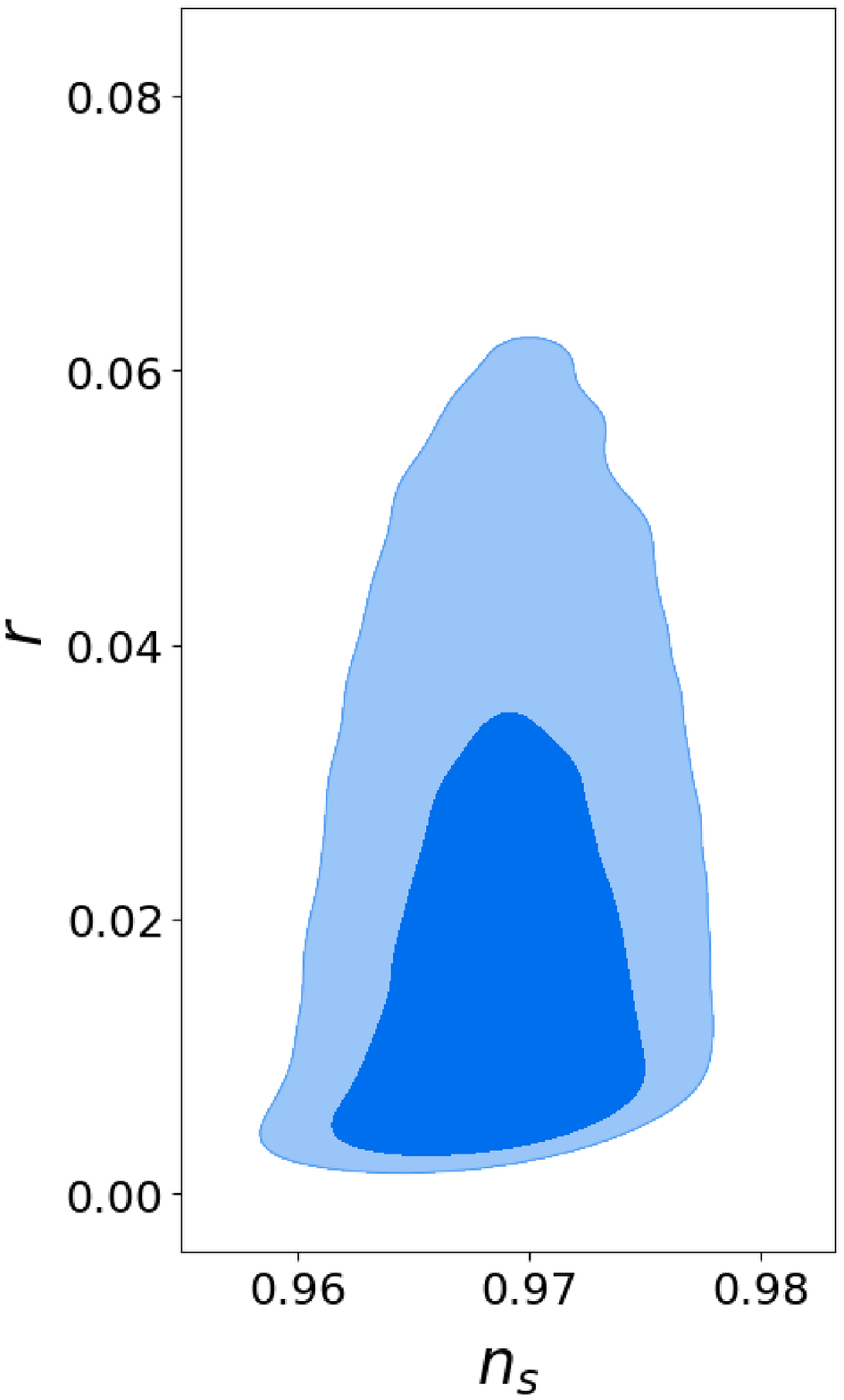}
\caption{Joint $68\%\, \, C.\, L.$, and $95\%\, \, C.\, L.$  constraints on $n_s$ and $r$ from Planck-2018, BICEP3 and 
BAO data}
\label{fig:nsr}
\end{center}
\end{minipage}
\end{figure}

The joint $68\%\, \, C.\, L. $ and $95\%\, \, C.\, L.$ 
constraints on the potential parameters $\beta$ and  $M$, and $N_{pivot}$ are shown in 
Fig.~\ref{fig:betamandn} and Fig.~\ref{fig:mandn}.
$\beta$ and $M$ from Planck-2018 and BICEP3 \cite{BICEP:2021xfz} data are presented in Fig. \ref{fig:betam}, which shows
that the two parameters are strongly correlated. The potential parameter $\beta$ is also strongly correlated with the number
of e-foldings $N_{pivot}$, as can be seen from the joint constraints on $\beta$ and $N_{pivot}$ in Fig. \ref{fig:betan}.
It is evident from the Fig. that  more the deviation from the Starobinsky model, the lesser e-foldings are preferred
by the Planck-2018 observations. Fig.~\ref{fig:mandn}  indicates that $N_{pivot}$ is also strongly correlated with the potential 
parameter $M$. 

The joint constraints on $r$ and $n_s$ are shown in Fig.~\ref{fig:nsr}. Here these two parameters are derived parameters and
the constraints  on these two parameters are derived from the constraints on the potential parameters and $N_{pivot}$, which
are used as an input parameters for MCMC analysis.

\section{Conclusions} \label{conclusions}
The $R^\beta$ term in Einstein Hilbert action  with $\beta$ slightly different from $2$ arises as a quantum correction to 
Starobinsky $R^2$ term \cite{Codello:2014sua,Ben-Dayan:2014isa,Rinaldi:2014gua}. 
Inflation with $R^\beta$ term, named as power law Starobinsky inflation, was first considered in 
\cite{Muller:1989rp,Gottlober:1992rg}. This model gained popularity in 2014 after BICEP2 reported large tensor-to-scalar 
ratio, and it was shown by \cite{Codello:2014sua,Martin:2014lra, Costa:2014lta} that large $r$ can be generated 
in $R^\beta$ inflation with $\beta$ slightly less than $2$. The analysis of power law Starobinsky inflation was
further done by \cite{Chakravarty:2014yda,Motohashi:2014tra,Odintsov:2022bpg} using slow-roll approximation, and constraints 
on the parameters of the potential (\ref{pot}), $\beta$ and $M$, were obtained from CMB constraints on 
inflationary parameters $n_s$ and $r$.  

In this work we analyze  power law Starobinsky inflation, in the light of Planck-2018 and BICEP3 \cite{BICEP:2021xfz} CMB observations and
other large scale structure observations. We use the inflaton potential (\ref{pot}) for power law Starobinsky inflation
in the Einstein frame. We  evaluate the power spectra for scalar and tensor perturbations
numerically using MODECODE. With the help of this we perform  MCMC analysis using COSMOMC to put constraints on the 
inflaton potential parameters $\beta$ and $M$, and the number of e-foldings $N_{pivot}$. We vary $\beta$ between
$1.9$ to $2.07$ to consider deviation from the Starobinsky inflation. We find from Planck-2018 and BICEP3 \cite{BICEP:2021xfz} observations that
$\beta = 1.966^{+0.035}_{-0.042},\, \, \, 95\%\, C.\, L.$. This implies that the current CMB and LSS observations
 prefer slight deviation from Starobinsky inflation. The value $\beta=2$ lies within $2\sigma$ of the best fit value. 
For our analysis we consider the general reheating scenario and we find
that the number of e-foldings from the end of inflation to the time when pivot scale $k_{pivot}$ leaves
the inflationary horizon  $N_{pivot} = 41\pm 10,\, \, \, 95\%,\, C.\, L.$. We also find that the number of e-foldings and 
the parameters $M$ and $\beta$ are strongly correlated Fig.~\ref{fig:betamandn} and Fig.~\ref{fig:mandn}. Planck-2018 data prefers
smaller $N_{pivot}$ for larger deviation from the Starobinsky inflation.  

Deviation from $\beta=2$ was also found in the analysis done by \cite{Motohashi:2014tra,Odintsov:2022bpg,Meza:2021xuq}
using slow-roll approximation.
In \cite{Motohashi:2014tra} selective values of $n_s$ and $r$, allowed by Planck-2013 data, were used to find values of
$\beta$ (denoted by $p$ there) and $N_{pivot}$, and it was shown that $1.9 \le\beta\le 2$. It is also
found in \cite{Motohashi:2014tra} that $(\beta, N_{pivot}) = (1.93,30)$ for $(n_s, r) = (0.96,0.05)$.
In \cite{Odintsov:2022bpg} Planck-2018 joint constraints on $n_s$ and $r$ were used to obtain the constraints on
$\beta$ and it was found that $1.9\le \beta\le 1.9999$ for $N_{pivot}=\left[50,60\right]$.
Our results shown in Table \ref{Tab:constraint} agree with the analysis of  \cite{Motohashi:2014tra,Odintsov:2022bpg},
however, we have performed a robust statistical analysis by exploring the entire allowable range for the parameters
 $\beta$, $M$ and $N_{pivot}$. With our approach we have obtained the best fit values for these parameters along with their 
marginalized probability distributions and joint constraints on them, which provides stronger statistical evidence for
$\beta$ lower than $2$. The value of $\beta$ obtained in \cite{Meza:2021xuq} is slightly larger than $2$ ($\beta=2.0008$), 
which deviates by $2\sigma$ from the best fit value (\ref{betadata})

It has been shown that the potential for the Starobinsky inflation in the Einstein frame can be obtained from the 
no-scale supergravity \cite{Ellis:2013xoa,Ellis:2013nxa,Ellis:2018zya}. The potential for power law Starobinsky inflation\
in the Einstein frame from no-scale supergravity is derived by \cite{Chakravarty:2014yda}. Since the variants of 
Starobinsky inflation can be obtained from supergravity, these models play an important role in particle physics
phenomenology. The bounds on inflaton potential parameters obtained in this work can be useful to build models of inflation
from supergravity that can help us in connecting inflation with other high energy physics phenomena. 

\section{ACKNOWLEDGEMENTS}
The authors would like to thank ISRO Department of Space Govt. of India to provide financial
support via RESPOND programme Grant No. DS\_2B-13012(2)/47/2018-Sec.II.

\end{document}